\documentclass[a4paper,11pt]{article}
\usepackage{color}
\usepackage{graphicx}
\usepackage[]{epsfig}
\usepackage{amsfonts}
\usepackage{amsmath}

\newcommand{\be}{\begin{equation}}
\newcommand{\ee}{\end{equation}}
\newcommand{\ba}{\begin{eqnarray}}
\newcommand{\ea}{\end{eqnarray}}

 \newcommand{\bea}{\begin{eqnarray}} \newcommand{\eea}{\end{eqnarray}}

\begin{document}
\title{\bf A Note on Holography and Phase Transitions}
\author{M.~Bellon$^{a,b}$, E.~F.~Moreno$^c$  and
F.~A.~Schaposnik$^d$\thanks{Associated with CICBA}
\\
\vspace{0.2 cm}
\\
 {\normalsize \it  $^{a}$UPMC Univ Paris 06, UMR 7589, LPTHE,
F-75005,
Paris, France }\\
{\normalsize \it $^b$ CNRS, UMR 7589, LPTHE, F-75005, Paris, France}
\\
{\normalsize \it $^c$Department of Physics, Northeastern University,
Boston, MA 02115, USA.} \\
{\normalsize \it $^d$\it Departamento de F\'\i sica, Universidad
Nacional de La Plata}\\ {\normalsize\it C.C. 67, 1900 La Plata,
Argentina}}
\date{\today}
\maketitle
\begin{abstract}
Focusing on the connection between the Landau theory of second
order phase transitions and  the holographic approach to
critical phenomena, we study diverse field theories in an
anti-de Sitter black hole background. Through simple analytical
approximations, solutions to the equations of motion can be
obtained in closed form which give rather good approximations
of the results obtained using more involved numerical methods.
The agreement we find stems from rather elementary
considerations on perturbation of Schr\"odinger equations.
\end{abstract}

\subsection*{Introduction}
Much activity has been centered in the last few years on the
application of the gauge/gravity duality, originally emerged
from string theory \cite{Maldacena}--\cite{Witten1998}, to
analyze strongly interacting field theories by mapping them to
classical gravity. More recently such duality was successfully
applied to describe, holographically, systems undergoing phase
transitions like superconductors, superfluids and other
strongly interacting systems \cite{Gubser1}--\cite{Lugo2} (for a wider list
of references see for example \cite{R1}).

In the holographic approach to the study of phase transitions
one starts, on the gravity side, with a field theory in
asymptotically Anti-de Sitter space-time with temperature
arising from a black hole metric, either introduced as a
background or resulting from back reaction of matter on the
geometry. Then, using the AdS/CFT correspondence, one can
study the behavior of the dual field theory defined on the
boundary, identify order parameters and analyze the phase
structure of the dual system.

Several models  with  scalar  and gauge fields in a bulk which
corresponds asymptotically to an AdS black hole metric have
been studied \cite{Gubser1}--\cite{Lugo2}.  The rather involved
systems of non-linear coupled differential equations that have
to be solved require in general application of numerical
methods. In this way one often finds nontrivial hairy
solutions that cease to exist for $T > T_c$, where $T$ is the
Hawking temperature associated to the black hole and $T_c$ a
critical temperature depending on the space-time dimensions and the parameters
of the model on the bulk.

The main point in this AdS/CFT based calculation is that the
asymptotics of the solution in the bulk encodes the behavior of the QFT at
finite temperature defined on the border. One finds
in general a typical scenario of second order phase transition.
The critical exponents can be computed with a rather good
precision and coincide with those obtained within the mean
field approximation in a great variety of models.

It is the purpose of this note to get an analytical insight
complementing the numerical results, with a focus on the
connection between the holographic approach and mean field
theory for the calculation of critical exponents. To this end,
it will be important to first stress some connections, already
signaled in \cite{Iqbal}, between the Landau  phenomenological
theory of second order phase-transitions and the gauge/gravity approach.
We will expose the importance of analyticity to explain the
similitude in the results for relevant physical quantities (critical exponents,
dependence of the critical temperature on the charge density and magnetic
field, etc.) in very diverse models.
This will be done first by stressing the  relationship between the equations of
motion and a Schr\"odinger problem, so that usual perturbative techniques
allow to prove the critical behavior. We then show that simple matching
conditions lead to results that broadly agree with elaborate numerical
calculations.

It should be noted that our calculations assume that the
backreaction of dynamical fields is negligible (probe
approximation) which is valid when the gauge coupling constant
is large. This approximation is useful to study the behavior
near the phase transition which is precisely the domain we will
analyze, comparing our analytical results with those obtained
numerically. In fact, the holographic results we have
previously obtained solving numerically the
Einstein-Yang-Mills-Higgs equations of motion both considering
the probe approximation \cite{Lugo1} and  the case in which
matter backreacts on the geometry \cite{Lugo2} show that when
parameters are such that condensation takes place the solutions
are very similar and the nature of the phase transition is
identical.

With this in mind and using an analytic approach
proposed in \cite{GKS}, we study the equations of motion for
different models, showing how one can determine in a very
simple way the critical behavior of the systems defined on the
border with a good agreement with numerical results.

\subsection*{Holography and mean field results}
In the Landau approach to second order phase transitions one
considers an order parameter $ {\cal O}$ and assumes
analyticity of  the free energy $F = F(T,  {\cal O}  )$,  which
can   be expanded in even powers of $  {\cal O} $
\be F(T,{\cal O}  ) = F=F_0 +  F_2[T]  {\cal O}^2+ F_4[T]
{\cal O}^4 + \ldots \ee
The dependence of the order parameter on $T$ is obtained by
minimizing $F$ as a function of $\cal O$. The next step is to
expand  the coefficients $F_2$ and $F_4$, in powers of $T-T_c$.
To ensure stability and a change of behavior at $T_c$, one
takes  $F_2= a(T-T_c)$ and $ F_4 = {b}/4$ with $a,b$ positive
constants. In this way, one finds that the minimum of $F$ is at
$ {\cal O}  = 0$ for $T>T_c$ (disordered phase) while for
$T<T_c$ one has
\be
{\cal O}  \sim (T_c - T)^{1/2}
\label{lan}
\ee

Let us give a brief description of gauge/gravity approach to
phase transitions to connect it to the Landau theory. On the
gravity side one considers a classical field theory in a
Schwarzschild--AdS black hole background (or one in which back
reaction of fields on an asymptotically AdS space leads to a
black hole solution). The choice of such a geometry is dictated
by the fact that the warped AdS geometry prevents massive
charged particles to be repelled to the boundary by a charged
horizon and as a result a condensate  floating  over the
horizon can be formed. To find such condensate one should look
for non trivial static solutions for the fields outside the
black hole by imposing appropriate boundary conditions. The
behavior of fields at infinity then allows to determine the
dependence of the order parameter on  temperature, identified
with the Hawking temperature of the black hole.

Basically, the free energy $F$ is identified with the minimum
of the action in the gravitational theory with prescribed
boundary condition for the fields. Regularity of the fields at
the horizon and smoothness of the background geometry, basic
assumptions within the gauge/gravity duality, imply
analyticity of this action. This is a first point of contact
with Landau theory of second order phase transitions leading to
a mean field behavior. We shall now see that if continuity and
smoothness of fields are imposed in the region between the
horizon and infinity, the mean field behavior found using a
numerical approach is reproduced with a good precision, as can
be seen  following a very simple analytic approach proposed in
\cite{GKS} which we apply below for different models.

\subsection*{The Abelian Higgs model  in $d$ space dimensions}

Dynamics of the system is governed by the action
\begin{equation}
S = \int d^{d+1}x \sqrt{|g|} \left( -\frac14 F_{\mu\nu} F^{\mu\nu} -
| \nabla_\mu - iA_\mu\Psi|^2 - m^2 |\Psi|^2 \right) \label{S}
\end{equation}
The background metric is the standard $AdS_{d+1}$-Schwarschild
black hole
\begin{eqnarray}
ds^2 &=& -f(r) dt^2 + \frac1{f(r)}  dr^2 + r^2 dx_idx^i
\; , \;\;\; i=1,2, \ldots,d-1\nonumber\\
f(r) &=& r^2\left(1 - \frac{r_h^d}{r^d}\right) \label{metrica}
\end{eqnarray}
with $i = 1,2,\ldots, d-1$ and we have chosen the AdS radius to
be unity. Different values for the mass term can be considered
\cite{HHH1}\cite{HR}: $m^2 = 0, -2$ and $m^2 = m^2_{BF} = -9/4$
in $d=3$ and $m^2 = 0, -3, -4$ in $d=4$, the last value
corresponding to the Breitenlhoner-Freedman bound $m^2_{BF}<0$
marking the boundary of stability for a scalar field in AdS
\cite{BF}. The black hole temperature is given by
\begin{equation}
T = \frac{d}{4\pi} r_h
\label{tempe}
\end{equation}

In order to look for simple classical solutions for this model
one can propose the following ansatz
\begin{equation}
\Psi = |\Psi| = \psi(r) \; , \;\;\;\; A_\mu = \phi(r) \delta_{\mu 0}
\label{ansatz}
\end{equation}
Imposing regularity of the solution at the horizon ($r=r_h$) one
gets
\begin{align}
&\psi(r_h) = \frac{d\,r_h}{m^2} \psi'(r_h)\cr
&\phi(r_h) = 0
\label{r=rh}
\end{align}
Concerning the asymptotic behavior of the scalar potential
$\phi$ and the scalar field $\psi$ one has
\begin{eqnarray}
\psi &=& \frac{\psi_-}{r^{\lambda_-}} +
\frac{\psi_+}{r^{\lambda_+}}\nonumber\\
\phi &=& \mu - \frac\rho{r^{d-2}} + \ldots
\label{boundary}
\end{eqnarray}
with
\begin{equation} \lambda_\pm = \frac12 \left( d \pm\sqrt{d^2 +
4m^2} \right)
\end{equation}
According to the gauge/gravity correspondence, $\mu$
corresponds to the chemical potential in the dual theory
defined on the boundary and $\rho$ to the charge density.
Concerning the scalar field $\psi$, both falloffs are
acceptable provided the following condition holds \cite{KB}
\be
-\frac{d^2}4 < m^2 < -\frac{d^2}4 + 1
\ee
One can then pick either $\psi_+ = 0$  or $\psi_- = 0$ leaving
a one parameter family of solutions so that the condensate of
the dual operator ${\cal O}^\sigma$ will be given by
\be \langle {\cal O}^\sigma \rangle = \psi_\sigma
\label{dictio}\ee
with  the boundary condition written as
\be
\varepsilon^{\sigma\rho} \psi_\rho = 0 \ee
where $ \sigma,\rho= +,-$ and $\varepsilon^{+-} = 1$. For
definiteness we shall impose the condition $\psi_-=0$

It will be convenient for the analysis that follows to change
variables according to
\begin{equation}
z = \frac{r_h}{r}
\label{cambio}
\end{equation}
so that the horizon is fixed at $z=1$ and the boundary at
$z=0$. In terms of this new variable $z$, the equations for the
functions in ansatz (\ref{ansatz}) become:
\begin{equation}
\psi'' -   \frac{d-1 + z^d}{z(1-z^{d})}\,
\psi' +  \left( \frac{\phi^2}{r_h^2(1-z^d)^2} - \frac{m^2}{z^2(1 -
z^d)}\right)\psi = 0 ,\label{system1}
\end{equation}
\begin{equation}\phi'' - \frac{d-3}{z}\phi' -
\frac{2\psi^2}{z^2(1 - z^d)} \phi = 0,
\label{system2}
\end{equation}
where the prime denotes $d/dz$.

Conditions (\ref{r=rh}) and (\ref{boundary}) now read
\begin{align}
\psi'(1) &= - (m^2/d)\, \psi(1)=0 \nonumber\\
\phi(1) &=0
\label{z=1}\\
\psi_B(z) &\simeq_{z\to 0} {D_+}\, {z^{\lambda_+}} + D_-\,
z^{\lambda_-}\nonumber\\
\phi_B(z) &\simeq_{z\to 0} \mu - q z^{d-2} \label{asymp}
\end{align}
with $ D_{\pm} = r_h^{-\lambda_\pm}\, \psi_\pm$ and
$q=\rho/r_h^{d-2}$.

In view of eq.(\ref{tempe}), the system
(\ref{system1})-(\ref{system2}) only depends on the black-hole
temperature $T$ through the non-linear $\phi^2 \psi$ term in
the scalar field equation (\ref{system1}): this describes the
coupling of the scalar to the electric potential and gives an
effective negative mass.

In the limit $T \to \infty$ the nonlinear term in
(\ref{system1}) vanishes so that the equation becomes linear
with no non-trivial soliton solutions. Now, as already pointed
out in \cite{Gubser2} the coupling of the scalar to the Maxwell
field is responsible for producing a negative effective mass
for $\psi$, and this effect becomes more important at low
temperatures. This indicates that one should expect an
instability taking place at some point towards forming scalar
hair.  At such point, the stabilizing effect of the $m$-term is
overcome. If a non-trivial solution exist at low temperatures,
with an asymptotic behavior associated with a non zero v.e.v.
of an order parameter $\langle O_\psi \rangle$  in the dual
theory, it is natural to expect that  a critical temperature
$T_c$ should exist such that $\langle O_\psi \rangle_{T>{T_c}}
= 0$. This was indeed verified numerically for different values
of the mass $m$ (including the Breitenlohner-Freedman bound
value) and various space-time dimensions
 \cite{HHH1},\cite{HR}.

Now, eq.~(\ref{system1}) for fixed potential $\phi$ can be
viewed as a Schr\"odinger equation with zero energy: in order
to have non trivial solutions, the resulting potential has to
adjust itself for the existence of a unique zero eigenvalue for
every temperature below the critical one. In the vicinity of
the critical point, the variation of the potential is linear in
the temperature variation, but quadratic in the normalization
of the field through the effect of the non linear term in
equation~(\ref{system2}). The eigenvalue of the
Schr\"odinger equation is, to first order, simply proportional
to some integral of the potential as can be seen from its
variational evaluation: with a suitable
weight function $a(z)$, the integral of $a(z)\psi(z)$ times the
left hand side of equation~(\ref{system1}) gives the eigenvalue for a
normalized $\psi$ and its variations are of second order with respect
to the variations of $\psi$. Fixing $\psi$ to the eigenfunction for $T=T_c$,
the two
sources of the variation of this integral through the potential must
compensate themselves and one obtains a \(\Delta T \propto
\phi^2\) relation. It is precisely this type of behavior that
has been found from the numerical solution to the  differential
equations on the gravity side.

To go further in the analysis without resorting to a numerical
analysis, we shall consider  expansions of the fields in the
bulk near $z=1$ and $z=0$.
Imposing the conditions of continuity and
smoothness of the solutions at a point $z_m$ intermediate
between the boundary ($z=0$) and the horizon ($z=1$) will give
algebraic equations between the parameters of the solution.

For the
solution near the horizon ($z=1$) we have, up to order
$(z-1)^2$:
\begin{align}
\psi_H(z)& = \psi_0 + \psi_1\, (z-1) + \frac{1}{2}\, \psi_2\,
(z-1)^2
\nonumber\\
\phi_H(z)& = \phi_0 + \phi_1\, (z-1) + \frac{1}{2}\, \phi_2\,
(z-1)^2 \label{expansion0}
\end{align}
with $\psi_0, \, \psi_1,\, \psi_2, \phi_0, \, \phi_1\, \phi_2$
constants. The boundary conditions (\ref{z=1}) at $z=1$
imply
\begin{align}
&\psi_1=-(m^2/d)\, \psi_0\\
&\phi_0=0
\end{align}
Substituting these values in \eqref{expansion0} and using the
differential equations (\ref{system1})-(\ref{system2})
we can obtain $\phi_2$ and $\psi_2$ as a function of $\phi_1$
and $\psi_0$. We get:
\begin{align}
\psi_H(z)& = \psi_0 + \frac{m^2}{d}\, \psi_0\, (1-z) +
\frac{2 d m^2 r_h^2+m^4 r_h^2 - \phi_1^2}{4 d^2 r_h^2}\, \psi_0\, (1-z)^2\\
\phi_H(z)& = -\phi_1\, (1-z) + \frac{1}{2}\,
\left(d-3-2\,{\psi_0^2}/{d}\right) \phi_1\, (1-z)^2
\end{align}

As announced, imposing the
conditions of continuity and smoothness at an intermediate
point $z_m$ allows to obtain a solution. Interestingly enough, as first
observed in
\cite{GKS} the result of this crude approximation are quite
stable with respect to the intermediate point, so we will
consider the case $z_m=1/2$. This can be understood from the
fact that $\psi$ is the ground state of a Schr\"odinger
equation, so that it cannot have nodes: additional terms in the
expansion of $\psi$ should rapidly fade away.

We will first analyze the boundary condition $D_-=0$, so we
have the set of equations
\begin{eqnarray}
\psi_H(1/2) = \psi_B(1/2)\;, & &
\psi'_H(1/2) = \psi'_B(1/2)\\
\phi_H(1/2) = \phi_B(1/2)\;, & &
\phi'_H(1/2) = \phi'_B(1/2)
\end{eqnarray}
to be solved for $\psi_0$, $\phi_1$, $D_+$, and $\mu$. We
obtain

\begin{align}
\psi_0^2 &=\frac{d}{2^{d+1}} \frac{16(d-2)\, B\, q + 2^d (d-5) \,r_h\, A}{r_h\,
A} \\
D_+ &= \frac{2^{\lambda_+-1} \, (4 d + m^2)}{d\, B^2}\, \psi_0\\
\phi_1 &=- \frac{r_h\, A}{B} \\
\mu &=\frac{1}{2^{d+2}} \frac{8\, d\, q\, B+ 2^d\, r_h\, A}{B}
\end{align}
(for simplicity $D_+$ is written in terms of $\psi_0$). Here
\begin{align}
A&= \sqrt{16\, d^2 \, \lambda_+ + m^4 (\lambda_+ + 2) + 2 \, d\, m^2 (6+5
\lambda_+)}\\
B&=\sqrt{\lambda_++2}
\end{align}
{Using the AdS/CFT dictionary \eqref{dictio}, we
can identify the v.e.v. $\langle \mathcal{O} \rangle $
 of the operator $\mathcal{O}$ dual to the scalar field with the
asymptotic coefficient $\psi_+$, $\langle \mathcal{O}\rangle
\equiv \psi_+ = r_h^{\lambda_+}\, D_+$.} Now, one can write
$D_+$ (or $\psi_0$ since they are proportional) as a function
of $ T = ({d}/{4\pi})\, r_h$. Remembering that $q=
{\rho}/{r_h^{d-2}}$ one has
\begin{equation}
\psi_0^2=\frac{d(5-d)}{2}\, \left(\frac{T_c}{T}\right)^{d-1} \left[
1- \left(\frac{T}{T_c}\right)^{d-1} \right]
\end{equation}
where we have defined
\begin{equation}
T_c^{d-1}=
\frac{ 2^{4-d} (d-2)}{(5-d)(4
\pi/d)^{d-1}}\frac{B}{A} \, \rho
\label{forae}
\end{equation}
We then have, for the order parameter
\be
\langle \mathcal{O}\rangle = C_d\,
\left(\frac{T}{T_c}\right)^{(\sqrt{d^2+4m^2}+1)/2}\,
\sqrt{1-\left(\frac{T}{T_c}\right)^{d-1}}
\label{crytical}
\end{equation}
with
\be
C_d = \frac{(4 d + m^2)}{(\lambda_+ + 2)}
\left(\frac{d}{2 \pi}\right)^{\lambda_+} \left(\frac{(5-d)}{8
d}\right)^{1/2} \label{cd} \ee

One can see from eq.(\ref{crytical}) that for $T$ close to
$T_c$ one has the typical second-order phase transition
behavior $\langle \mathcal{O} \rangle \propto \sqrt{1-T/T_c}$.
Note that $T_c \propto \rho^{1/(d-1)}$ in agreement with the
change of dimensions of the charge density for different $d$'s.

Our results for the critical temperature, as inferred from
(\ref{forae})-(\ref{cd}) in the cases  $d=3,4$, $m^2=-2$ (with
$z_m = 1/2$) are $T_c=0.15 \rho^{1/2}$ and  $T_c=0.2
\rho^{1/3}$ respectively. They can be compared with those
obtained using a different analytical approximation based on
perturbation theory near the critical temperature. For the case
$m = m_{BF}$, we obtain $T_c=0.12 \rho^{1/2}$ and  $T_c=0.25
\rho^{1/3}$\cite{ST}--\cite{STM} in very good agreement with the
exact numerical results, $T_c=0.15 \rho^{1/2}$ and $T_c=0.25
\rho^{1/3}$. One can conclude that there is a good quantitative
agreement between the three sets, which is not much affected by the
choice of the point $z_m$ in the method. This
last fact was already observed in \cite{GKS} for the particular
case $d=3$  compared with the numerical results given
in \cite{HHH1}.

\subsection*{Non Abelian gauge field  in $d=3$ space dimensions}

We now consider the case of an $SU(2)$ Yang-Mills theory in
$3+1$ dimensional Anti-de Sitter--Schwarzschild background, as
a prototype for the gauge/gravity duality in the case of pure
gauge theories \cite{G}--\cite{Pufu}. The action is
\begin{equation}
S = -\frac14 \int d^{4}x \sqrt{|g|}
F^a_{\mu\nu} F^{a\,\mu\nu}
\end{equation}
with the field strength defined as
\begin{equation}
F^a_{\mu\nu} = \partial_\mu A^a_\nu - \partial_\nu A^a_\mu +
g \epsilon^{abc} A^b_\mu A^c_\nu \; , \;\;\;\; a=1,2,3
\end{equation}
Writing the gauge field as an isospin vector
\be
\vec A = (A^1,A^2,A^3) = (A_\mu^a  dx^\mu)
\ee
we consider the following ansatz  for solving the equations of
motion
\begin{equation}
\vec A =      J(r)  dt \, {\check e}_3 -  K(r)    \rho d\varphi\,
{\check e}_\rho +   K(r)  d\rho \,{\check e}_\varphi \label{ansatzz}
\end{equation}
where $r$ is the radial variable in spherical coordinates and
\begin{eqnarray}
x &=& \rho \cos \varphi \nonumber\\
y &=& \rho \sin \varphi
\end{eqnarray}
The background metric   is
\begin{equation}
ds^2 =  - f(r)  d^2 t + \frac1{f(r)}  d^2 r + r^2 (dx^2 + dy^2)
\end{equation}
with
\begin{equation} f(r) =   r^2\left(1 -
\frac{r_h^3}{  r^3}\right) \label{background}
\end{equation}
In terms of the $z$ variable defined in (\ref{cambio})  the equations
of motion read
\begin{align}
\left((1-z^3)K' \right)'&= \frac{1}{r_h^2} \left(K^2 - \frac{J^2z^2}{1 - z^3}
\right)K\nonumber\\
J''(z) &= \frac{2}{r_h^2}\, \frac{J(z) K^2(z)}{1-z^3}
\label{ecomo}
\end{align}
%

At the horizon ($z=1$), $J$ must vanish, so we have the
following expansions for the fields, up to order $(z-1)^2$,
\begin{align}
K_H(z)& = K_0 + K_1\, (z-1) + \frac{1}{2}\, K_2\, (z-1)^2
\nonumber\\
J_H(z)& = J_1\, (z-1) + \frac{1}{2}\, J_2\, (z-1)^2
\label{expansion1}
\end{align}
Using the equations of motion~(\ref{ecomo}), the coefficients $K_1$, $K_2$
and $J_2$ can be written in terms of $K_0$ and $J_1$,
\begin{align}
K_H(z)& = K_0 - \frac{K_0^3}{3 r_h^2}\, (z-1) +
\frac{-J_1^2 K_0 r_h^2+ 6 K_0^3 r_h^2+ 3 K_0^5}{36 r_h^4}\, (z-1)^2\\
J_H(z)& = J_1\, (z-1) - \frac{J_1 K_0^2}{3 r_h^2}\,(z-1)^2
\end{align}

At the $z=0$ boundary  one has the asymptotic expansions
\begin{align}
K_B(z) &= C_1/r_h\, z \label{asympK}\\
J_B(z) &= D_0 - D_1/r_h z \label{asymp1}
\end{align}
where $D_1$ can be associated with the charge density.
Coefficient $C_1$ in (\ref{asympK}) should be identified with
the order parameter $\langle {\cal O}_K\rangle$ for the theory
on the border. Since the order parameter is related to a vector
field ($A_i$), the associated theory on the border is a p-wave
superconductor \cite{Pufu}.

As in the scalar field case, we will match both solutions at an
intermediate point which we again choose as $z_m=1/2$,
\begin{eqnarray}
K_H(1/2) = K_B(1/2) \; , & &
K'_H(1/2) = K'_B(1/2) \nonumber\\
J_H(1/2) = J_B(1/2) \; , & &
J'_H(1/2) = J'_B(1/2)
\label{mathching}
\end{eqnarray}
and solve these equations for $K_0$, $J_1$, $C_1$, and $D_0$.
From the first two identities \eqref{mathching}, we obtain:
\begin{align}
C_1&=\frac{4}{3} r_h\, K_0 + \frac{K_0^3}{9 r_h}\\
J_1&=-\sqrt{48 r_h^2+22 K_0^2 + 3 K_0^4/r_h^2}
\end{align}
(we chose $J_1<0$ so $J(z)>0$).

Substituting these values in eq.\eqref{mathching}, we can
obtain $K_0$ as the root of the following polynomial in
$K_0^2$:
\begin{equation}
\frac{3}{r_h^2}\, K_0^8+ 40\, K_0^6 + 207 r_h^2\, K_0^4 + 486
r_h^4\, K_0^2- 9 r_h^2 (D_1^2-48 r_h^4)=0
\end{equation}

This equation implies that $K_0^2$ vanishes as
\begin{equation}
K_0^2 = \frac{D_1^2-48 r_h^4}{54 r_h^2} + O(D_1^2-48 r_h^4)^2
\end{equation}

Finally, introducing the temperature $T=3/(4 \pi)\, r_h$, we
can write
\begin{equation}
K_0^2 = \frac{128 \pi^2}{81}\, \frac{T_c^4}{T^2}
\left[1-\left(\frac{T_c}{T} \right)^4
\right]\; \;\;\;\; \text{for $T$ near $T_c$}
\end{equation}
where
\begin{equation}
T_c^2= \frac{3 \sqrt{3}}{64\, \pi^2} D_1
\label{anterior}
\end{equation}
Then, for $T$ close to $T_c$ we have the typical second-order
phase transition behavior for $C_1 =\langle {\cal O}_K\rangle
\propto \sqrt{1-T/T_c}$, in good agreement with
refs.\cite{G}--\cite{Pufu}. Our numerical value for $T_c$ from
equation \eqref{anterior}, $T_c/\sqrt{D_1} = 0.091$ can be
compared with the numerical value obtained in \cite{Pufu},
$T_c/\sqrt{D_1} = 0.125$.

\subsection*{The $d=3$ scalar case in the presence of
an applied  magnetic field}

We shall consider here a system with dynamics governed by the
action (\ref{S}) in the $d=3$, $m^2 = -2 $ case, when an
external magnetic field $H$ is applied. This corresponds to the
case where the background metric is an $AdS_{3+1}$ {\it
magnetically} charged black hole \cite{Nakano}--\cite{AlJohn}.
\begin{eqnarray}
ds^2 &=& -f(r) dt^2 + \frac1{f(r)} dr^2 + r^2 dx_idx^i \; ,
\;\;\; i=1,2\nonumber\\
f(r) &=&  \frac{r^2}{L^2} - \frac{M}{r} + \frac{H^2}{r^2}
\label{metricaH}
\end{eqnarray}
The $ f(r)= 0 $ condition for having event horizons gives a
quartic algebraic equation with 4 roots which can be explicitly
written in terms of surds. There are two complex conjugate
roots which we shall call $r_1$ and $r_2$ and two real roots
$r_3< r_4$ We shall then call   $r_h \equiv r_4$ the external
black hole horizon. Taking $L=1$ from here on, one has
\be r_h = \frac{1}{2\, 6^{1/3}} \left(B + \sqrt{\frac{12 M}{B}
- B^2}\right) \ee
where
\[ B= \sqrt{\frac{8\, 3^{1/3} H^2}{(D+9M^2)^{1/3}}+ 2^{1/3}
(D+9M^2)^{1/3}} \;\; , \;\;\; D=\sqrt{81 M^4-768 H^6}
\]
The actual form of the other roots $r_a$ ($a=1,2,3$) is not
necessary since one will only need the standard relationships
between roots and coefficients
\bea
&& r_1 + r_2 + r_3 = - r_h \nonumber\\
&& r_1r_2 +r_1r_3 + r_2r_3 = r_h^2 \nonumber\\
&& r_1r_2r_3 = H^2/r_h
\eea
together with the relation
\begin{equation}
\frac{H^2}{r_h^4} - \frac{M}{r_h^3} + 1 = 0
\end{equation}
The Hawking temperature associated to the black hole is
\begin{equation}
T = \frac{f'(r_h)}{4\pi}=\frac{1}{4\pi} \left(3 r_h -
\frac{H^2}{r_h^3}\right)
\label{tempe2}
\end{equation}

In terms of the variable $z=r_h/r$ the equations of motion read
\begin{align}
&\tilde \psi''(z)+\frac{d f(r_h/z)/dz}{f(r_h/z)} \tilde \psi'(z)
+ \frac{r_h^2}{z^4}
\left(\frac{\tilde \phi^2(z)}{f(r_h/z)^2}+\frac{2}{f(r_h/z)}\right)=0
\nonumber\\
&\tilde \phi''(z)- 2\frac{\psi^2(z)}{f(r_h/z)}\, \tilde \phi(z)
\left(\frac{r_h^2}{z^4}\right)=0
\label{llaH}
\end{align}
where $\tilde \psi(z)\equiv \psi(r_h/z)$ and $\tilde \phi(z)
\equiv \phi(r_h/z)$ (but from now on, the tilde will be
omitted).
In terms of the new variables we have:
\begin{align}
&f(r_h/z) = (r_h/z)^2 (1-z)(1-\tilde r_1 z) (1-\tilde r_2 z) (1-\tilde r_3 z)\\
&\frac{d f(r_h/z)/dz}{f(r_h/z)} = \frac{-2 - M z^3/r_h^3 + 2 H^2 z^4/r_h^4}
{z (1-z)(1-\tilde r_1 z) (1-\tilde r_2 z) (1-\tilde r_3 z)}
\end{align}
where $\tilde r_a=r_a/r_h\, , \;\; a=1,2,3$.

The boundary conditions for system (\ref{llaH}) at the horizon
are
\begin{align}
&\phi(1)=0 \\
&\psi'(1) = \frac{2 r_h}{3 r_h - H^2/r_h^3} \psi(1)
\label{asynh}
\end{align}
while asymptotically one has
\begin{align}
&\psi(z) = D_1 z + D_2 z^2 \\
& \phi(z) = \mu - (\rho/r_h) z
\end{align}

For the solution of system (\ref{llaH}) near the horizon  we
have the expansions, up to order $(z-1)^2$:
\begin{align}
\psi_H(z)& = \psi_0 + \psi_1\, (z-1) + \frac{1}{2}\, \psi_2\, (z-1)^2
\nonumber\\
\phi_H(z)& = \phi_0 + \phi_1\, (z-1) + \frac{1}{2}\, \phi_2\, (z-1)^2
\label{expansion3}
\end{align}
with $\psi_0$,  $\psi_1$, $\psi_2$, $\phi_0$, $\phi_1$ and $\phi_2$
constants. Using \eqref{asynh} we get
\begin{align}
&\phi_0=0\\
&\psi_1 =\frac{2 r_h}{3 r_h - H^2/r_h^3} \psi_0
\end{align}
Substituting these values in \eqref{expansion3} and using the
differential equations we can obtain $\phi_2$ and $\psi_2$ as
functions of $\phi_1$ and $\psi_0$. We get:
\begin{align}
&\psi(z) = \psi_0 + \frac{2 r_h^4}{H^2 - 3 r_h^4}\, \psi_0\, (1-z)-
\frac{r_h^4(8 r_h^4 + r_h^2 \phi_1^2 - 12 H^2)}{4(H^2 - 3 r_h^4)^2}\,
\psi_0\, (1-z)^2
\\
&\phi(z) = -\phi_1 (1-z) + \frac{r_h^4\, \psi_0^2\, \phi_1}{H^2 - 3 r_h^4}
(1-z)^2
\end{align}

As before, we impose matching conditions at $z=1/2$:
\begin{eqnarray}
\psi_H(1/2) = \psi_B(1/2) \; , &&
\psi'_H(1/2) = \psi'_B(1/2)\\
\phi_H(1/2) = \phi_B(1/2) \; , & &
\phi'_H(1/2) = \phi'_B(1/2)
\end{eqnarray}
and solve for $\psi_0$, $\phi_1$, $D_2$, and $\mu$. We look for
solutions with $D_1=0$ and $D_2$ corresponding to the order
parameter of the $2+1$ system defined on the boundary. We
obtain:
\begin{align}
&\phi_1=-2 R(r_h,H) \\
&\mu=\frac{H^2 \left({\rho }/{r_h}+2 R(r_h,H)\right)-r_h^4
\left({3 \rho }/{r_h} + R(r_h,H) \left(\psi_0^2 +
6\right)\right)}{2 \left(H^2-3 r_h^4\right)}\\
&\psi_0^2 = \frac{\left(3 r_h^4-H^2\right)
\left( {\rho }/{r_h}-2
R(r_h,H)\right)}{2 r_h ^4  R(r_h,H)}\\
& D_2 =\frac{\left(88 {r_h}^8- {\phi_1}^2 {r_h}^6-68 H^2 {r_h}^4
+ 16 H^4\right) \psi_0}{4 \left(H^2-3
{r_h}^4\right)^2}
\label{matched}
\end{align}
where
\be
R(r_h,H) = \sqrt{\frac{7 {r_h}^8-6 H^2 {r_h}^4+2 H^4}{{r_h}^6}}
\ee

The equation for $\psi_0^2$  in terms of the dimensionless
variable $u = r_h/\sqrt{H}$ takes the form \be \psi_0^2
=\frac{(3u^4 -1) \left(\rho/H - 2\sqrt{2/u^4 -6 +7u^4} \right)}
{2u^4\sqrt{2/u^4 -6 +7u^4}} \label{enr1} \ee with the
temperature (\ref{tempe2}) given by \be \frac{T}{\sqrt H} =
\frac1{4\pi}\left(3u - \frac1{u^3}\right) \label{enr2} \ee

The minimum value that $u$  can take is the one for which
$T=0$, $u = 3^{-1/4}$, and corresponds to the condition
$3^3M^4= 2^7H^6$. From this, we see that in order to have a
non-trivial solution the following inequality should hold \be H
\leq \frac{\rho}{ 2\sqrt{2/u^4 -6 +7u^4}} \ee The maximum of
the r.h.s. is attained for $u_m = (2/7)^{1/8}>u_0$ so that
there is a critical value $H_c$ of the magnetic field beyond
which no non-trivial solution exists,
\be H_c = 0.41 \rho. \ee

\begin{figure}[h]
\epsfxsize=3. in
\begin{center}
\leavevmode
\epsffile{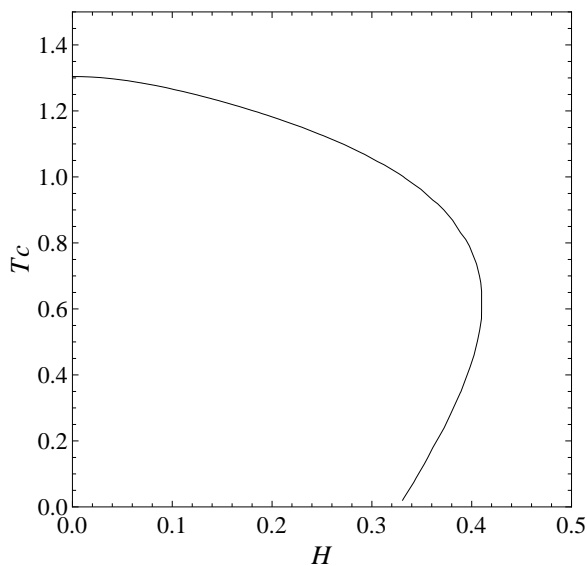}
\end{center}
\caption{The phase diagram of $T_c$ against the
magnetic field $H$. The condensed phase ($D_2 \ne 0$)
corresponds to the lower left part below the line. The critical
temperature decreases as the magnetic field grow up to the
critical value $H_c$.}
\label{figure 1}
\end{figure}

\begin{figure}[h]
\epsfxsize=3.5 in
\begin{center}
\leavevmode
\epsffile{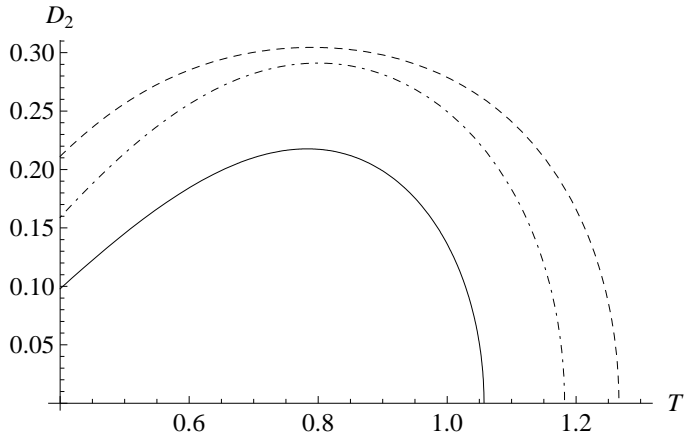}
\end{center}
\caption{A plot of the order parameter $D_2$ as a
function of temperature $T$. The charge density is $\rho=1$.
The dashed line corresponds to a magnetic field $H=0.1$, the
dash-dot one to $H=0.2$ and the solid one to $H=0.3$. (the
critical value is $H_c=0.6$).}
\label{figure 2}
\end{figure}

Using eqs. (\ref{enr1})-(\ref{enr2}) one can determine the
critical temperature $T_c$ as a function of $H$. We give in
Figure \ref{figure 1} the resulting $T_c = T_c(H)$ curve.
Interestingly, we find that in the range \be H_c> H>
\frac{\rho}{ 2\sqrt{2/u_0^4 -6 +7u_0^4}} = 0.327 \rho \ee the
curve $T_c(H)$ becomes double valued so that a nontrivial
solution only exists in the range $T_{c_2} >T>T_{c_1}$
%
%
According to the gauge/gravity duality $D_2$ should be
identified with the order parameter, $\langle {\cal
O_\psi}\rangle = D_2 $. We obtain the following expression
\be
D_2 = \frac{5u^4-2}{6u^4-2}
\frac{\sqrt{\rho/H -
2\sqrt{2/u^4 -6 +7u^4}}}{u^2(2/u^4 - 6 +7u^4)^{1/4}}
\ee
Note that $D_2$ becomes negative for $u_0<u<(2/5)^{1/4}$ or,
equivalently, for $0<T<0.0316\sqrt{H}$. However, when $H\to 0$
the numerator and denominator coefficients of $D_2$ that
multiply $\psi_0$ cancel out. This is consistent with the
result of the no-magnetic field model. (In fact, we have
checked that the results of both models,  $H=0$, $d=3$ and
$m^2=-2$,  are identical.  We present in Figure \ref{figure 2}
curves for $D_2$ as a function of temperature for different
values of the external magnetic field.

\vspace{0.3 cm}

The results are qualitatively in agreement with those described
in \cite{Nakano}--\cite{AlJohn} where the reported curves
obtained by numerically solving the equations of motion are
similar to those in Figure 2.

\subsection*{Summary and discussion}

We have analyzed a number of models which have been proposed to
study phase transition through the AdS/CFT correspondence. The
common feature of all three models we discussed was that the
space time bulk geometry was an Anti-de Sitter black hole.
Although the dynamical field content was very different ---a
charged scalar coupled to an electric potential, the same model
in an external magnetic field and a pure non-abelian gauge
theory--- the emerging scenarios are very similar and always
include a second order phase transition with mean-field
critical exponents.

On general grounds, we were able to explain why the highly
symmetric ans\"atze generally used, produce the critical
behaviors seen in mean field theory or the Landau approach.
Founded on basic principles as the connection between the
equations of motion and the Schr\"odinger equation, we clarify
the similarity between several relevant quantities along a
variety of models. In particular we showed that resorting to
simple matching conditions we obtain closed form solutions that
significantly agree with the results obtained by numerically
solving the exact set of equations of motion. This uncovers the
important role played by analyticity to explain the universal
behavior of certain physical constants.

The method seems to work very well near the critical
temperature, though it deviates from the numerical results as
we approach $T\to 0$. In this regime our approach should be
refined.

Alternative  analytic calculations have been recently presented
in \cite{HNu} and \cite{ANu} where the phase transition
vicinity is studied solving the equations of motion in terms of
a series expansion near the horizon. Although the approach in
these works is close to the one proposed in \cite{GKS} and
applied here, the possibility in the latter of varying the
intermediate point $z_m$ at which the matching is performed
allows to obtain better solutions at fixed order $N$ in the
expansion, as already pointed out in \cite{ANu}.  As we have
seen, in the matching approach the problem reduces to find the
solution of an algebraic equations system and this can be done,
to the order we worked here, in a straightforward way.
Increasing the order will of course complicate the algebraic
system but in view of its main features, it can be handled by a
simple computational software like Mathematica, at least for
the next few orders. For a large-order expansion the method
followed in refs.\cite{HNu} and \cite{ANu} seems to be more
appropriate.

Although the matching method works very well near the critical
temperature, it deviates from the numerical results as $T$
approaches to 0. In this regime the method should be refined.
In particular it is to be expected that taking into account the
quantum fluctuations of the gravity theory one should be able
to go beyond mean field approximation results. Also, one should
consider generalized Lagrangians (like the St\"uckelberg one
considered in \cite{Franco}) leading to various types of phase
transitions (first or second order with both mean and non-mean
field behavior) as parameters are changed. There is also the
possibility that including fermions in the bulk model could
substantially change the critical behavior of the theory in the
bulk (see \cite{McG} and references therein). The simplicity of
the approach presented here, not requiring refined numerical
calculations,   should be an asset when trying to explore these
more complex situations.



\end{document}